\documentclass[10pt]{article}

\usepackage{amsmath}
\usepackage{amssymb}
\usepackage{graphics}
\usepackage{rotating}
\usepackage{cite}
\usepackage{color}
\usepackage{fancybox}

\def\0{\mbox{\tiny $0$}}
\def\1{\mbox{\tiny $1$}}
\def\2{\mbox{\tiny $2$}}
\def\3{\mbox{\tiny $3$}}
\def\4{\mbox{\tiny $4$}}
\def\5{\mbox{\tiny $5$}}
\def\6{\mbox{\tiny $6$}}
\def\7{\mbox{\tiny $7$}}
\def\8{\mbox{\tiny $8$}}
\def\9{\mbox{\tiny $9$}}
\def\L{\mbox{\tiny $L$}}

\def\F{\mbox{\tiny $F$}}
\title{\shadowbox{\large \bf
GRAPHENE TESTS OF KLEIN PHENOMENA}}

\author{
\small  Stefano De Leo\thanks{Department of Applied Mathematics,
State University of Campinas, Brazil [deleo@ime.unicamp.br] } \,\,
and\, Pietro Rotelli\thanks{Department of Physics, University of
Salento and INFN Lecce, Italy [rotelli@le.infn.it]}}

\date{\small
\fcolorbox{black}{yellow} {\color{red} $\bullet$ {\color{black}{
{\footnotesize  {\sc Journal of Physics A} {\bf 44}, 475305-9 (2011)}}} {\color{red}{$\bullet$}} } }

\begin{document}
%
%%%%%%%%%%%%%%%%%%%%%%%%%%%%%%%% PAPER %%%%%%%%%%%%%%%%%%%%%%%%%%%%%%%%%%%%%

\maketitle

\vspace*{-.7cm}

\begin{abstract}
\noindent Graphene is characterized by chiral electronic
excitations. As such it provides a perfect testing ground for the
production of Klein pairs (electron/holes). If confirmed, the
standard results for barrier phenomena must be reconsidered with,
as a byproduct, the accumulation within the barrier of holes.
\end{abstract}

%%%%%%%%%%%%%%%%%%%%%%%%%%%%%%%%%%%%%%%%%%%%%%%%%%%%%%%%%%%%%%%%%%%%%%%
%%%%%%%%%%%%%%%%%%%%%%%%%%%%%%%%%%%%%%%%%%%%%%%%%%%%%%%%%%%%%%%%%%%%%%%

%%%%%%%%%%%%%%%%%%%%%%%%%%%%%%%%%%%%%%%%%%%%%%%%%%%%%%%%%%%%%%%%%%%%%%%
%%%%%%%%%%%%%%%%%%%%%%%%%%%%%%%%%%%%%%%%%%%%%%%%%%%%%%%%%%%%%%%%%%%%%%%

% Graphene electronic transport, 72.80.Vp
% Graphene films, 68.65.Pq

% Warning: No PACS code given

%02.10.Hh Rings and algebras
%02.10.Ud Linear algebra
%02.10.Yn Matrix theory

%02.30.Hq Ordinary differential equations
%02.30.Jr Partial differential equations
%02.30.Tb Operator theory

%03.65.-w Quantum mechanics
%03.65.Ca Formalism
%03.65.Ta Foundations of quantum mechanics;
%03.65.Xp Tunnelling, traversal time, quantum Zeno dynamics

%12.15.F Quarks and lepton masses and mixing
%14.60.Pq Neutrino mass and mixing

%\offprints{~Stefano De Leo.}

%%%%%%%%%%%%%%%%%%%%%%%%%%%%%%%%%%%%%%%%%%%%%%%%%%%%%%%%%%%%%%%%%%%%%%%
%%%%%%%%%%%%%%%%%%%%%%%%%%%%%%  SECTION   %%%%%%%%%%%%%%%%%%%%%%%%%%%%%
%%%%%%%%%%%%%%%%%%%%%%%%%%%%%%%%%%%%%%%%%%%%%%%%%%%%%%%%%%%%%%%%%%%%%%

\section*{\normalsize I. INTRODUCTION}

Graphene is a subject of intense interest in the scientific
community. After the
 assignment of the 2010 Nobel prize in Physics, for the experimental realization of
 graphene obtained by extracting single atomic thick crystallites
 from bulk graphite,  to Geim and
 Novoselov\cite{NOS1,NOS2}, many papers appear every day
with new and interesting discussions about the properties of  this
intriguing material. Due to this vast amount of literature
periodic updates are needed to cover the recent progress. The most
explored aspects of graphene physics are related to its electronic
properties. Due to the fact that electrons propagating through
graphene lose their effective mass, they are described by a
Dirac-like equation rather than the Schrodinger equation. This
open the doors to new possibilities to explore effects that were
previously inaccessible. The massless Dirac equation in graphite
intercalation compounds was first pointed out in 1984\cite{DVM}.
Nevertheless, the first observation of the Dirac fermion nature of
electrons in grephene\cite{FO,DC} was only provided  when graphene
was mechanically exfoliated on Si0$_{\2}$\cite{NOS1,NOS2}.
Relativistic phenomena related to Zitterbewegung\cite{KAT1},
chirality and minimal conductivity\cite{KAT2,KAT3}, Klein
tunneling\cite{KAT4}, quantum Hall effect\cite{HE1,HE2}, Schwinger
production\cite{SCH}, Casimir interactions\cite{CAS},
transformations of discrete Landau level spectrum to a continuum
of extended states in presence of a static electric
field\cite{LAN}  have been investigated and when possible tested
by using this reach and surprising material. For an extensive theoretical
review of the properties of electron transport in graphene, we refer
the reader to the report of Castro\cite{NOS3}.

In the last year, as observed by Geim\cite{GEIM}, the initial
skepticism with respect to graphene applications was gradually
evaporating and graphene has rapidly changed its status from newcomer to
star.  The study presented in this paper was inspired by the possibility to understand,
by an experiment with graphene, the physics which governs a well known phenomenon of the Dirac equation known as  Klein paradox\cite{KLE}, i.e. for the step potential the reflected flux can be larger than the incident
one\cite{ZUB}. This relativistic effect is attributed to the fact that  sufficiently strong potentials
open the door to pair production and consequently to positrons inside the potential region. This result is generally accepted for a step potential but not always quoted for a barrier potential where plane wave solutions {\em seem} not to exhibit pair production\cite{DEL4}. Pair production at the discontinuity of square potentials has never been observed experimentally because its observation requires very strong electric fields\cite{GRIB}.
The transmission of  electron through graphene heterojunctions\cite{KS1,KS2,KS3} could now provide as an experimental signature of the Klein pair production.

Graphene is a two dimensional structure characterized by chiral
(mass zero) electronic exitations one of whose equation of motion
reads
\begin{equation}
-\,i\,\hbar\,v_{\F}\,\boldsymbol{\sigma}\cdot \nabla\,
\psi_{\F}(\boldsymbol{r})=E\,\psi_{\F}(\boldsymbol{r})\,\,,
\end{equation}
where $\boldsymbol{\sigma}=(\sigma_{\1},\sigma_{\2},0)$ and
$v_{\F}\approx 10^{^{6}}\,\mbox{m/sec}$ is the Fermi velocity.
This equation mimics one of the 2-dimensional free Weyl
equations\cite{WEY,ZUB} for a (Weyl) spinor if we substitute
$v_{\F}$ by the velocity of light $c$. In the following, we shall
present the results for the Weyl equation for a potential barrier
via the well tried step method\cite{DEL1,DEL2,DEL3}. Consequently,
we ``reconsider'' the hypothesis of Klein pair
production\cite{KLE} for the step potential. The step potential is
actually a purely abstract construct, because of its infinite
extension (same abstraction of plane waves). However, as we have
demonstrated elsewhere with the Dirac equation\cite{DEL2}, the
step potential is a useful tool for general piece-wise potential
calculations. Contrary to a common misconception, it is compatible
with a very large barrier potential. We will have the opportunity
to reassert this fact here with our results for the Weyl equation.
Furthermore, and this is essential to our conclusions, if Klein
pair production occurs for a step, it must {\em also} occur for a
barrier potential\cite{DEL4}.

For the Dirac equation\cite{DIR,ZUB}, the so-called Klein energy
zone (where Klein pair production is possible) requires a step
potential height $V_{\0}$ of at least $2\,m$ ($m$ the mass of the
fermion). For a free  electron this means $V_{\0}>1\,\mbox{MeV}$.
Not a typical laboratory potential energy step. Not surprisingly,
experimental verification of Klein pair production has not been
achieved to date. Neutrinos are a possible alternative choice of
particle since they are now known to have a very small  mass, but
the weak nature of their interactions introduces formidable
experimental difficulties\cite{NEU}. The best alternative
``particle'' is that of electrons within  a dielectric with
effective masses $m_*$ much smaller  than the free electron mass.
Graphene, with a null effective electron mass, is the most
promising material of all to study. A step potential of any height
could test the creation of Klein pairs (electron-hole) and as we
have anticipated this has consequences even  for the barrier
potential seen within the graphene structure.

In the next section, we calculate, in 3-dimensions for
completeness, the chiral interaction with a step and subsequently
in section III that with a barrier potential. The arguments
require a discussion of an incoming wave packet and resolves the
apparent paradox that a step has a single transmitted wave packet
(both for the diffusion zone and the Klein zone) while a long
barrier gives rise to infinite wave packets. We shall again argue
that nevertheless the step and barrier results are perfectly
compatible. In section IV, we discuss Klein pair production. We
also show that for this case the transmission/reflection amplitude
series are divergent. The finite standard (matrix) results are
radically different. They do not involve pair production. The
application to graphene is straightforward. We argue that the
barriers within graphene act as a ``well-trap'' for the created
``antiparticles''. This should result in a growth in time of
positive charge within the well/barrier. We suggest the detection
of such an accumulating charge as proof of Klein pair production.

\section*{\normalsize II. WEYL INTERACTION WITH A STEP POTENTIAL}

The spinor Weyl equation for a positive helicity (massless) state
is\cite{ZUB}
\begin{equation}
-\,i\,\boldsymbol{\sigma}\cdot \nabla\,
\Psi(\boldsymbol{r},t)=i\,\partial_t\,\Psi(\boldsymbol{r},t)\,/\,c\,\,.
\end{equation}
Plane wave solutions of the kind $\psi(\boldsymbol{r})=
u(\boldsymbol{p})\,\exp[\,i\,(\,\boldsymbol{p}\cdot\boldsymbol{r}-Et)/\hbar]$,
where $E=|\boldsymbol{p}\,|\,c$ and $\boldsymbol{p}$ is the
particle three-momentum, result in the spinor equation
\begin{equation}
\boldsymbol{\sigma}\cdot \boldsymbol{p}\,\,
u(\boldsymbol{p})=E\,u(\boldsymbol{p})\,/\,c\,\,,
\end{equation}
the solution of which is
\begin{equation}
u(\boldsymbol{p})=\sqrt{\frac{E+p_{\3}c}{2\,E}}\,\,\,\left[\,  1
\,\,\,\,\,\,\,
\frac{p_{\1}c+i\,p_{\2}c}{E+p_{\3}c}\,\right]^{t}\,\,,
\end{equation}
where the upper index $t$ represents the transpose. Our step
potential is chosen in the $x$ variable. Without loss of
generality, we set the discontinuity at $x=0$,
\[
V(x)=\left\{\,0\,\,\,\mbox{for}\,\,x<0\,\,\,\mbox{region
I},\,\,\,\,\,V_{\0}>0\,\,\,\mbox{for}\,\,x>0\,\,\,\mbox{region
II}\,\right\}\,\,.
\]
 In region I ($x<0$), we have set the incoming
momentum as $\boldsymbol{p}=(p_{\1},p_{\2},p_{\3})$ with
\[p_{\1}c=\sqrt{E^{^{2}}-(p_{\2}c)^{^{2}}-(p_{\3}c)^{^{2}}}\,\,.\]
In region II ($x>0$), only $p_{\1}$ changes. We call the new
momentum $\boldsymbol{q}=(q_{\1},p_{\2},p_{\3})$ with
\[q_{\1}c=\sqrt{(E-V_{\0})^{^{2}}-(p_{\2}c)^{^{2}}-(p_{\3}c)^{^{2}}}\,\,.\]
The continuity equation for $\Psi(\boldsymbol{r},t)$ at $x=0$ is
\begin{equation}
\left[\,  1 \,\,\,\,\,\,\,
\frac{p_{\1}c+i\,p_{\2}c}{E+p_{\3}c}\,\right]^{t}+
r_{\0}\,\left[\,  1 \,\,\,\,\,\,\,
\frac{-p_{\1}c+i\,p_{\2}c}{E+p_{\3}c}\,\right]^{t}=
\underbrace{\sqrt{\frac{E\,(E-V_{\0}+p_{\3}c)}{(E-V_{\0})\,
(E+p_{\3}c)}}}_{N}\,\,\,t_{\0}\,\left[\,
1 \,\,\,\,\,\,\,
\frac{q_{\1}c+i\,p_{\2}c}{E-V_{\0}+p_{\3}c}\,\right]^{t}
\end{equation}
 After a straightforward calculation, we find
\begin{equation}
r_{\0} = \frac{1-\alpha}{1+\alpha}\,\,\,\,\,\,\,\,\mbox{and}\,\,
\,\,\,\,\,\, t_{\0}=\frac{2}{N\,(1+\alpha)}\,\,,
\,\,\,\,\,\,\,\,\,\,\mbox{with}\,\,\,\,\,
\alpha=\frac{q_{\1}(E+p_{\3}c) +
i\,p_{\2}V_{\0}}{p_{\1}(E-V_{\0}+p_{\3}c)}\,\,.
\end{equation}
Starting from the reflection probability,
\begin{equation}
|\,r_{\0}|^{^{2}}=\frac{1+|\alpha|^{\2}-2\,\mbox{Re}[\alpha]}{1+|\alpha|^{\2}+
2\,\mbox{Re}[\alpha]}\,\,,
\end{equation}
there are specific cases to be considered.

The situation when $q_{\1}^{\2}<0$, i.e. when
$p_{\2}^{\2}+p_{\3}^{\2}>[(E-V_{\0})/c]^{^{2}}$,  implies an
imaginary $q_{\1}$ and whence to ``tunneling'' when penetration
into the classically forbidden region II occurs. An imaginary
momentum $q_{\1}$ also implies $\mbox{Re}[\alpha]=0$, consequently
$|r_{\0}|=1$. We are not interested in evanescent solutions in
this paper and henceforth consider only real momentum $q_{\1}$.

Diffusion occurs when
$E>V_{\0}+\sqrt{p_{\2}^{\2}+p_{\3}^{\2}}\,c$. In this case
$\mbox{Re}[\alpha]>0$ and consequently $|r_{\0}|<1$. We treat this
case in the next section. For
$E<V_{\0}-\sqrt{p_{\2}^{\2}+p_{\3}^{\2}}\,c$, we have again a real
momentum $q_{\1}$ but now
$E-V_{\0}+p_{\3}c<E-V_{\0}+\sqrt{p_{\2}^{\2}+p_{\3}^{\2}}\,c<0$
and consequently $\mbox{Re}[\alpha]<0$ which implies $|r_{\0}|>1$.
This {\em Klein} energy zone is thus characterized by an
oscillatory (free) plane wave in region II. The corresponding flux
could, in principle, be interpreted as an {\em additional}
incoming particle beam from $x=+\infty$ hence resulting in an
excess of ``reflected'' particles in region I, or in alternative
as an antiparticle flux flowing in the positive $x$ direction and
``created'' at $x=0$. The former choice contradicts our initial
conditions of a sole incoming particle wave from the left
($x=-\infty$) it also implies the unpalatable concept of free {\em
particles} that exist and propagate freely  in the classically
forbidden region below potential. The latter solution implies the
creation of {\em Klein pairs}. The ``antiparticles'' (energy $-E$)
travel to the right while the particles created (energy $E$) add
to the reflected incoming particles and travel to the left, thus
yielding $|r_{\0}|>1$. This latter interpretation is consistent
with the fact that antiparticles see a potential of $-V_{\0}$ and
hence are above potential {\em free antiparticles}, since
$-E>-V_{\0}$. However, it must be recalled that this latter
interpretation also implies that particle/antiparticle number is
not conserved (only {\em total} charge and other additive quantum
numbers are conserved). Our spinor equation, as occurs for the
Dirac equation, would thus no longer be a single particle
equation. Some have claimed this to be another positive feature of
the Dirac equation since it anticipated pair creation in field
theory.

\section*{\normalsize III. WEYL DIUFFUSION WITH A BARRIER POTENTIAL}

The barrier potential $V(x)= \{ V_{\0}\,\,\,
\mbox{for}\,\,0<x<L\,\,,\,\,\, 0 \,\,\,\mbox{elsewhere}\}$ can be
treated for calculational purposes as a two ``step'' interaction.
As we have shown in previous publications\cite{DEL1,DEL2,DEL3} it
is also physically correct to consider it as such. A brief
argument for this is as follows. Consider a numerical simulation
of an incoming wave packet from the left, say with
$E>V_{\0}+\sqrt{p_{\2}^{\2}+p_{\3}^{\2}}\,c$ so that diffusion
occurs. Upon reaching $x=0$ (the barrier) the wave packet splits
into two amplitudes. One the reflected wave, the other the
transmitted wave moving in the region of the potential. The group
velocity of the latter is of course lower than the
incoming/reflected wave group velocity. When the wave packet
reaches the end of the barrier, reflection and transmission occurs
anew. The very existence of a reflected wave for a downward step
is characteristic of quantum mechanics. This procedure repeats
itself ad infinitum. Infinite reflected and transmitted wave
packets are produced. If $L$ is large compared to the wave packet
size, the individual wave packets will be separated and
interference effects are negligible. Incoherence reigns. Even when
the barrier size is of order or smaller than the wave packets this
approach can be used except that now the overlapping wave packets
amplitudes are coherent and must be summed. Separation into
individual contributions may even be done for single plane waves
which, of course, are totally  coherent. The sum of the infinite
contributions for plane waves yields exactly the same result as
that deduced from a matrix calculation of the continuity
equations. This argument can therefore be reversed. A wave packet
convolution of a (matrix) barrier solution will yield infinite
outgoing wave packets when $L$ grows larger than the wave packet
size. This is by no means obvious, but numerical calculations
confirm this claim\cite{DEL1,DEL2}.

We have called the above method of calculation the two step
approach to the barrier. However, it actually requires three step
results since the discontinuity at $x=0$ receives waves impinging
both from the left (the original wave) and from the right
(reflected waves at $x=L$). These three results are listed below.
Any two can be derived from the third by implementing the momentum
changes indicated, together with the
appropriate plane wave phase changes.\\

\noindent $\bullet$ {\sc Step 1} [$x=0$] (see section II for derivation),\\
incoming momentum $(E/c,p_{\1},p_{\2},p_{\3})$, reflected
$(E/c,-p_{\1},p_{\2},p_{\3})$, transmitted
$[(E-V_{\0})/c,q_{\1},p_{\2},p_{\3}]$,
\begin{equation}
\begin{array}{lcl}
r_{\0}&=&\displaystyle{\frac{1-\alpha}{1+\alpha}}\,\,,\\ \\
t_{\0}&=&\displaystyle{\frac{2}{N\,(1+\alpha)}}\,\,,
\end{array}
\,\,\,\,\,\,\,\,\mbox{with}\,\,\, \alpha=\frac{q_{\1}(E+p_{\3}c) +
i\,p_{\2}V_{\0}}{p_{\1}(E-V_{\0}+p_{\3}c)} \,\,.
\end{equation}
\\

\noindent $\bullet$ {\sc Step 2} [$x=L$],\\ incoming
momentum $[(E-V_{\0})/c,q_{\1},p_{\2},p_{\3}]$, reflected
 $[(E-V_{\0})/c,-q_{\1},p_{\2},p_{\3}]$,
  transmitted $(E/c,p_{\1},p_{\2},p_{\3})$,
\begin{equation}
\begin{array}{lcl}
r_{\L}&=&\displaystyle{\frac{1-\beta}{1+\beta}\,\,\,e^{2\,i\,q_{\1}L/\hbar\,}}\,\,,\\
\\
t_{\L}&=&\displaystyle{\frac{2\,N}{1+\beta}\,\,\,
e^{i\,(q_{\1}-p_{\1})L/\hbar}}\,\,,
\end{array}
\,\,\,\,\,\,\,\mbox{with}
\,\,\,\beta=\frac{p_{\1}(E-V_{\0}+p_{\3}c)-ip_{\2}V_{\0}}{q_{\1}(E+p_{\3}c)}\,\,.
\end{equation}
\\

 \noindent $\bullet$ {\sc Step 3} [$x=0$],\\ incoming
momentum $[(E-V_{\0})/c,-q_{\1},p_{\2},p_{\3}]$, reflected
$[(E-V_{\0})/c,q_{\1},p_{\2},p_{\3}]$,
 transmitted $(E/c,- p_{\1},p_{\2},p_{\3})$,
\begin{equation}
\begin{array}{lcl}
\widetilde{r}_{\0}&=&\displaystyle{\frac{1-\gamma}{1+\gamma}}\,\,,\\
\\ \widetilde{t}_{\0}&=&\displaystyle{\frac{2\,N}{1+\gamma}}\,\,,
\end{array}
\,\,\,\,\,\,\,\mbox{with}\,\,\,
\gamma=\frac{-p_{\1}(E-V_{\0}+p_{\3}c)-ip_{\2}V_{\0}}{-q_{\1}(E+p_{\3}c)}=\beta^{^*}\,\,.
\end{equation}
\\

\noindent Observing that
$1+\beta^{^*}=(1+\alpha)/\mbox{Re}[\alpha]$ and
$1-\beta=(\alpha-1)/\mbox{Re}[\alpha]$, we can rewrite all the
previous amplitudes in terms of $\alpha$,
\begin{equation}
\begin{array}{lclclclclcl}
r_{\0}&=&\displaystyle{\frac{1-\alpha}{1+\alpha}}\,\,, & &
r_{\L}&=&\displaystyle{\frac{\alpha
-1}{1+\alpha^*}\,\,\,e^{2\,i\,q_{\1}L/\hbar}}\,\,, & &
\widetilde{r}_{\0}&=&\displaystyle{\frac{\alpha^*-1}{1+\alpha}}\,\,,\\
\\t_{\0}&=&\displaystyle{\frac{2}{N\,(1+\alpha)}}\,\,, & &
t_{\L}&=&\displaystyle{\frac{2\,N\,\mbox{Re}[\alpha]}{1+\alpha^*}\,\,\,
e^{i\,(q_{\1}-p_{\1})L/\hbar}}\,\,, & &
\widetilde{t}_{\0}&=&\displaystyle{\frac{2\,N\,\mbox{Re}[\alpha]}{1+\alpha}}\,\,.
\end{array}
\end{equation}
 The total transmitted wave beyond the barrier ($x>L$) is then
given by
\begin{equation*}
t  =   t_{\0}t_{\L}\,\sum_{s=\,0}^{\infty}
(r_{\L}\widetilde{r}_{\0})^{^s}=
\frac{t_{\0}t_{\L}}{1-r_{\L}\widetilde{r}_{\0}} =
4\,\mbox{Re}[\alpha]\,\,\frac{e^{-ip_{_1}L/\hbar}}{e^{-iq_{_1}L/\hbar}|1+\alpha|^{^2}-
e^{iq_{_1}L/\hbar}|1-\alpha|^{^2}}\,\,.
\end{equation*}
which incidently is normalization independent. After simple
algebraic manipulations, we find
\begin{equation}
\label{tra}
 t= e^{-ip_{\1}L/\hbar}\,\mbox{\Huge
/}\,\left[\cos(q_{\1} L/\hbar)-
i\,\frac{1+|\alpha|^{^2}}{2\,\mbox{Re}[\alpha]}\,\,\sin(q_{\1}
L/\hbar)\right]\,\,.
\end{equation}
Similarly, for the total reflected amplitude, we have
\begin{equation*}
r  =  r_{\0}+  t_{\0}r_{\L}
\widetilde{t}_{\0}\,\sum_{s=\,0}^{\infty}
(r_{\L}\widetilde{r}_{\0})^{^s}=r_{\0}+
\frac{r_{\L}\widetilde{t}_{\0}}{t_{\L}}\,t =
\frac{1-\alpha}{1+\alpha}\,\left[\,1 -
e^{i\,(p_{\1}+q_{\1})\,L/\hbar}\,t\,\right] \,\,,
\end{equation*}
which results in
\begin{equation}
\label{ref}
r=-\,i\,\frac{(1-\alpha)(1+\alpha^{*})}{2\,\mbox{Re}[\alpha]}
\sin(q_{\1} L/\hbar)\,\mbox{\Huge /}\,\left[\cos(q_{\1} L/\hbar)-
i\,\frac{1+|\alpha|^{^2}}{2\,\mbox{Re}[\alpha]}\,\,\sin(q_{\1}
L/\hbar)\right]\,\,.
\end{equation}
For diffusion, the single plane wave reflection and transmission
probabilities are $|\,r\,|^{^{2}}$ and $|\,t\,|^{^{2}}$ and of
course satisfy
\begin{equation}
|\,r\,|^{^{2}}+\,\,|\,t\,|^{^{2}}=\,1\,\,.
\end{equation}
The above results coincide with those calculated by solving the
coupled continuity equations. However, it must be recalled,
because often forgotten that even for diffusion $|t|^{^{2}}$ is
the transmission probability (i.e. has a physical meaning) {\em
only} for a single plane wave. For wave packets small compared to
the barrier width incoherence dominates. The transmission
probability is then an infinite sum of squares and {\em not} the
square of an infinite sum. In the limit of {\em total incoherence}
the probabilities become
\begin{eqnarray}
\label{prob}
\mbox{Incoherent T-Probability} & : &
|\,t_{\0}t_{\L}|^{^{2}}\,\sum_{s=\,0}^{\infty}
|\,r_{\L}\widetilde{r}_{\0}|^{^{2\,s}} =
\left(\,1-|\,r_{\0}|^{^{2}}\right)^{^{2}} \,\sum_{s=\,0}^{\infty}
|\,r_{\0}|^{^{4\,s}}=\frac{1-|\,r_{\0}|^{^{2}}}{1+|\,r_{\0}|^{^{^{\,2}}}}\,\,,
 \nonumber \\ \\
\mbox{Incoherent R-Probability}  & : & |\,r_{\0}|^{^{2}} +
|\,t_{\0}r_{\L}\widetilde{t}_{\0}|^{^{2}}\,\sum_{s=\,0}^{\infty}
|\,r_{\L}\widetilde{r}_{\0}|^{^{2\,s}} =|\,r_{\0}|^{^{2}} +
|\,r_{\0}|^{^{2}}\,
\frac{1-|\,r_{\0}|^{^{2}}}{1+|\,r_{\0}|^{^{^{\,2}}}}
=\frac{2\,|\,r_{\0}|^{^{2}}}{1+|\,r_{\0}|^{^{^{\,2}}}}\,\,,
\nonumber
\end{eqnarray}
with again probability conservation.

 Now the apparent conundrum, the step potential has only a
single reflected and transmitted wave. How can this be reconciled
with the infinite contributions (for large $L$)described above?
The answer is very simple. {\em The first contributions at $x=0$
are, obviously, the step results}. The others require a time for
their appearance which are multiples of $L/v_g$, where $v_g$ is
the group velocity of the wave packet in the barrier region. As
$L\to \infty$ this time interval goes to infinity. Formally, even
for the step the other contributions exist but one must attend an
infinite time interval to see even the second contribution and so
forth. This feature is ``hidden'' when using (unconvoluted) plane
waves since no barrier size exceeds the size of a plane wave.

An interesting observation is that the standard resonance
condition, $q_{\1}L=n\pi\hbar$, is only valid in the limit of total
coherence. However, another resonance condition is brought to
light in the above results. The total transmission amplitude
equals unity when there is ``head on'' diffusion,
$p_{\2}=p_{\3}=0$. This resonance condition does not depend upon
the incoming energy value. Furthermore, it is independent of
coherence requirements. The result occurs {\em step by step} for
head on collision. This latter resonance condition is a
consequence of zero mass. It does not appear for a massive Dirac
particle.

%\noindent $\bullet$ $\bullet$ $\bullet$ $\bullet$ $\bullet$\\
%Resonances $|T|=1$:\,\,\,\,\, $, q_{\1}L=n\pi$ only valid in the
%wave limit\,\,\,,\,\,\,\,\,$p_{\2}=p_{\3}=0\,[\,\alpha=\beta=1\,]$
%always valid
%\begin{eqnarray*}
%q_{\1}L/\hbar = n \pi & \,\,\,\to \,\,\, & T =
%(-1)^{^{n}}\,\exp[-ip_{\1}L/\hbar\,]\,\,\,\,\,\mbox{and}\,\,\,\,\,
%R=\frac{1-\alpha}{1+\alpha}\,+ \frac{1-\beta}{1+\beta^{*}}=0\\
%p_{\2}=p_{\3}=0 & \,\,\,\to \,\,\, & T =
%\exp[(ip_{\1}-q_{\1})\,L/\hbar\,]\,\,\,\,\,\mbox{and}\,\,\,\,\,
%R=0
%\end{eqnarray*}
%\noindent $\bullet$ $\bullet$ $\bullet$ $\bullet$ $\bullet$\\

\section*{\normalsize IV. KLEIN PAIR PRODUCTION}

In the previous section, we used the two step method for the
calculation of the total transmission and reflection amplitudes.
Both can be expressed as infinite sums of amplitudes. Each term in
the sum represents a wave packet source (once convolution is
performed). The series are convergent. The total amplitudes can be
calculated either by the two step method or by the  matrix method
based upon continuity at both $x=0$ and $x=L$. Which one uses is a
matter of taste.

Note that there is always oscillatory behavior for the $y$ and $z$
axis. It is only the $x$-axis (that of potential discontinuity)
where $q_{\1}$ may be either real (oscillatory behavior) or
imaginary (evanescent behavior). Bypassing, as previously
anticipated,  the evanescent case (tunneling)  which exists even
for mass zero particles when there is a transverse momentum
component, we now pass to the Klein energy region,
$E<V_{\0}-\sqrt{p_{\2}^{\2}+p_{\3}^{\2}}\,c$.

We have argued for the step result $|r_{\0}|>1$. This means that
the first contribution to the total reflection amplitude for a
barrier exceeds unity. This is already incompatible with the
matrix solution which always satisfies $|r|<1$

The situation is even more radical. The holes (antiparticles)
produced at $x=0$ have energy $-E$. They travel within a well
potential $V(x)=-V_{\0}$ for $0<x<L$ and zero beyond. At $x=L$
they are within their own Klein zone. Thus, they also create Klein
pairs, holes/electrons. The holes are reflected to the left.
Consequently, all holes are {\em entrapped} within the
barrier/well region. Pair creation occurs at each potential
discontinuity and hence with time the hole density within the
barrier/well increases.

The series for $t$ and $r$ contain the loop factor
\[ r_{\L}\widetilde{r}_{\0} = \frac{1+\,|\alpha|^{^{2}}-2\,\mbox{Re}[\alpha]}{
1+\,|\alpha|^{^{2}}+2\,\mbox{Re}[\alpha]}\,\,\,e^{2\,i\,q_{\1}L/\hbar}\,\,.\]
The energy zone for diffusion,
\[
E>V_{\0}+\sqrt{p_{\2}^{\2}+p_{\3}^{\2}}\,c\,\,\,\,\,
\Longrightarrow \,\,\,\,\, E>V_{\0}- p_{\3}\,c \,\,\,\,\,
\Longrightarrow \,\,\,\,\,\mbox{Re}[\alpha]>0\,\,\,\,\,
\Longrightarrow \,\,\,\,\,|r_{\L}\widetilde{r}_{\0}|<1\,\,.
\]
In this case, the series for $t$ and $r$ can be summed, see
Eq.(\ref{tra}) and (\ref{ref}). However, in the Klein energy zone,
\[
E<V_{\0}-\sqrt{p_{\2}^{\2}+p_{\3}^{\2}}\,c\,\,\,\,\,
\Longrightarrow \,\,\,\,\, E<V_{\0}- p_{\3}\,c \,\,\,\,\,
\Longrightarrow \,\,\,\,\,\mbox{Re}[\alpha]<0 \,\,\,\,\,
\Longrightarrow \,\,\,\,\,|r_{\L}\widetilde{r}_{\0}|>1 \,\,.
\]
Thus, the two step series in the Klein energy zone cannot be
summed. The matrix method is here {\em incompatible} with the two
step calculation. Holes entrapped within the potential well
(barrier) will bounce back and forth an infinitum increasing at
each reflection their number. Thus, the number of pairs created
grows without limit in time. This is of course only theoretical
since we expect the corresponding growth in (antiparticle) charge
to eventually modify the potential itself.

\section*{\normalsize IV. CONCLUSIONS}

In the previous sections, we have presented the results of the
three dimensional interaction of a chiral (Weyl) fermion with
first a step and consequently (using the step results) with a
barrier. We have considered separately the case of diffusion,
$E>V_{\0}+\sqrt{p_{\2}^{\2}+p_{\3}^{\2}}\,c$, and Klein zone,
$E<V_{\0}-\sqrt{p_{\2}^{\2}+p_{\3}^{\2}}\,c$. We have bypassed
tunneling in this work.

Now for graphene, we must set $p_{\3}=0$ so that our solutions
apply to a physical plane $(x,y)$. We must also allow for the
Fermi velocity by the substitution, $c\to v_{\F}$. In this limit
\[\frac{1+|\alpha|^{^2}}{2\,\mbox{Re}[\alpha]} =
\frac{E(E-V_{\0})/c^{\2}-p_{\2}^{\2}-p_{\3}^{\2}}{p_{\1}q_{\1}}\,\,
\longrightarrow\,\,\frac{E(E-V_{\0})/v_{\F}^{\2}-p_{\2}^{\2}}{p_{\1}q_{\1}}\,\,.
\]
 It is
convenient to introduce the angles $\phi=\arctan(p_{\2}/p_{\1})$
and $\theta=\arctan(p_{\2}/q_{\1})$. In terms of these angles,
\[
\left(\frac{E}{p_{\1}v_{\F}}\,,\,\frac{E-V_{\0}}{q_{\1}v_{\F}}\,,\,
\frac{p_{\2}^{\2}}{p_{\1}q_{\1}}\right)= \left(
\frac{\mbox{sign}[E]}{\cos\phi}\,,\,
\frac{\mbox{sign}[E-V_{\0}]}{\cos\theta}\,,\,\tan\phi \,\tan\theta
\right)\,\,.
\]
Thus, from Eq.(\ref{tra}) we find  the formulas standard to
graphene litterature\cite{NOS3,KAT4},
\begin{equation}
\label{gf}
 |\,t\,|^{^{2}}=1\,\mbox{\Huge
/}\,\left[\cos^{\2}(q_{\1} L/\hbar)+\left(
\frac{\mbox{sign}[E]\,\mbox{sign}[E-V_{\0}]-\sin\phi\,\sin\theta}{\cos\phi\,
\cos\theta}\right)^{^{2}}\,\,\sin^{\2}(q_{\1} L/\hbar)\right]\,\,.
\end{equation}
This equation clearly shows that beyond the standard resonances, $q_{\1} L =n\pi\hbar$,
there are additional head-on resonances for chiral fermions which are absent for massive particles.\\

We now recall some of our results.\\

\noindent
$\bullet$ For diffusion, in the limit of total coherence (wave packets large if compared to the barrier width)
the transmission probability is given by Eq.(\ref{tra}). For wave packets small compared to the barrier width, incoherence dominates and the transmission probability is given by Eq.(\ref{prob}). The first term in the infinite sum reproduces the step result.\\

\noindent
$\bullet$ As for the Klein zone, we have reiterated  our belief in Klein
pair production. Obligatory, and generally accepted for a step
potential, but not always the result quoted for a barrier
potential. Our claim is that this is the consequence of unwittingly
summing a divergent series. The coherent diffusion amplitudes {\em cannot}
be extended to the Klein energy zone.\\

\noindent
$\bullet$ The holes created via pair production are trapped within the potential region and
these localized holes have a continuous  energy spectrum.\\

The question of Klein pair production is one more hypothesis that
graphene could help to test. If valid it implies a very different
physics for ``above potential'' (Dirac/Weyl diffusion) and ``below
potential'' (Klein diffusion) phenomena. As always we leave final
judgement to experiment.

\section*{\small \rm ACKNOWLEDGEMENTS}

The authors thank anonymous referees for drawing the attention to interesting papers,
for their constructive comments and  for their useful suggestions. One of the authors (SdL) also
thanks the Department of Physics, University of Salento (Lecce, Italy), for the hospitality and the
FAPESP (Brazil) for financial support by the Grant No. 10/02213-3.

\end{document}